\documentclass[prd,showpacs,nofootinbib,preprintnumbers,a4,twocolumn]{revtex4}
\usepackage{amsmath}
\usepackage{amsfonts}
\usepackage{graphicx}
\usepackage{dcolumn}
\usepackage{hyperref}
\usepackage{amssymb}

\begin{document}
\title{On the Gauss-Bonnet Gravity}
\author{Naresh Dadhich}
\email{nkd@iucaa.ernet.in}
\affiliation{IUCAA, Post Bag 4,
Ganeshkhind, Pune 411 007, INDIA}
\begin{abstract}
We argue that propagation of gravitational field in the extra dimension
is motivated by physical realization of second iteration of self
interaction of gravity and it is described by the Gauss-Bonnet term.
The most remarkable feature of the Gauss-Bonnet gravity is that
at high energy it radically transforms radial dependence from
inverse to proportionality as singularity is
approached and thereby making it weak. Similar change over also occurs in 
approach to singularity in loop quantum gravity. It is analogous to Planck's
law of radiation where similar change occurs for high and low energy
behavior. This is how it seems to
anticipate in qualitative terms and in the right sense the
quantum gravity effect in $5$ dimensions where it is physically
non-trivial. The really interesting question is, could this
desirable feature be brought down to the $4-$dimensional
spacetime by dilatonic coupling to the Gauss-Bonnet term or
otherwise?
\end{abstract}

\maketitle

The most distinguishing feature of gravitation is that it is
universal and hence links to everything that physically exists
including massive as well as massless particles and above all
with itself. Its linkage to massless particles can only be
negotiated through curved space \cite{n1}. That means gravity
must curve spacetime and its dynamics has thus to be entirely
determined by the spacetime curvature, the Riemann curvature
tensor. We have no freedom to make any prescription, Newton's law
should follow from the Riemann curvature. It indeed does through
the Bianchi differential identity satisfied by the curvature
tensor.It leads to the Einstein equation which contains the
Newton's law in the limit \cite{n1}. \\
The Einstein equation so obtained naturally contains the so called
cosmological constant $\Lambda$ in a natural way as a constant of integration
without any reference to cosmology. It comes on the same footing as the
matter tensor in the equation and is indeed a new constant of the theory. It
really indicates the distinguishing feature of gravitation that here
spacetime background was not fixed as was the case for the rest of physics but
was dynamic describing gravitational force. $\Lambda$ is the measure of this
property. The Einstein equation is valid in all dimensions where Riemann
curvature is
defined,i.e. $n\ge2$. It is well-known that in dimensions $<4$, it is not
possible to realize the free field dynamics. Thus we come to the
usual $4$-dimensional spacetime. That means $4$ dimensions are necessary for
description of gravitational field. The question is, are they sufficient
too? \\
Let us now turn to the property of self interaction of gravitation which can
be evaluated only by an iteration process \cite{n2}. \footnote{ There is a 
long and distinguished history of deriving the Einstein equation from the 
Newtonian gravity through perturbative inclusion of self interaction (see 
\cite{dp}). However the self interaction we are referring here is the 
inherent property of the Einsteinian gravity and which is evaluated in the 
dynamic curved spacetime framework.}  
The spacetime metric is potential for the Einstein equation which contains
its second
derivative and square of the first derivative. It thus contains the first
order iteration through the square of the first derivative. The natural
question that arises is, how do we stop at the first iteration? We should go
to second and higher orders as well.
The basic entity at our disposal is the Riemann curvature, so should we square
it and add to the usual Einstein-Hilbert action of the Ricci scalar? This will
also square the second derivative which is the highest order of derivative.
If the highest order of derivative does not occur linearly in an equation,
then there will be more than one equation, and the question of having
unique solution does not arise. The property that highest order of derivative
occurs linearly is known as quasi-linearity. Is it then possible to have
higher
powers of first derivative yet the second derivative remaining linear? Yes,
the differential geometry offers a particular combination,
known as the Gauss-Bonnet (GB) given by
$R_{abcd}R^{abcd} - 4R_{ab}R^{ab} + R^2$,
which ensures the quasi-linearity character of the resulting equation. This
particular combination cancels out the square of the second
derivative. However it turns
out that this term makes no contribution in the equation for dimension $<5$.
We are thus forced to go to the extra $5$th dimension for the physical
realization of second iteration of self interaction of gravity. This is an
important conclusion we have reached simply by hooking onto the iterative
realization of self interaction. \\
Now the question arises, where does this iteration process of going to
higher dimension stop? If all the matter fields are confined to
$3-$brane/space, the $5-$dimensional bulk is completely free of matter and
hence
it is homogeneous and isotropic in space and homogeneous in time, and thereby 
maximally symmetric. It is 
therefore of constant curvature, an Einstein space with vanishing Weyl
curvature.
That means there is no more free gravity to propagate any further in the
higher dimension. The iteration chain thus naturally terminates at the second
iteration in $5-$dimensional bulk for matter fields living on the
$3-$brane. In how many dimensions should matter live has however to be
determined by the dynamics of matter fields. \\
The gravitational dynamics in the $5-$dimensional bulk \cite{d1,d2} is
described by
\begin{eqnarray} \label{gb}
G_{AB} = \alpha H_{AB} - \Lambda g_{AB}
\end{eqnarray}
where $G_{AB} = R_{AB} - \frac{1}{2} Rg_{AB}$, and
\begin{eqnarray}
H_{AB} &=& -2\left(RR_{AB}-2R_{AC}R^C_B -2R^{CD}R_{ACBD} \right. \nonumber \\
        &&\left. +R_A^{CDE}R_{BCDE} \right) +\frac{1}{2}g_{AB}\left(R^2 \right.\nonumber\\
       &&\left. -4R_{CD}R^{CD} +R_{CDEF}R^{CDEF}\right).\nonumber \\
\end{eqnarray}
Here $\alpha$ is the parameter coupling the Einstein-Hilbert
action with the GB term. It is easy to see that the condition of
constant curvature solves this equation to give an Einstein space
with redefined $\Lambda$ given by 
\begin{equation}
\lambda = \frac{3}{2\alpha}[-1\pm \sqrt{1 + \frac{4}{3}\alpha\Lambda}]
\end{equation}
which in the first approximation reduces to $\Lambda$, ~~ $-\Lambda - 
\frac{3}{\alpha}<0$ for $\alpha>0$. The former with +ve sign has the 
$\alpha\to0$ limit 
leading to the Einstein case. It is flat when $\Lambda=0$, which means GB
contribution has no independent existence. It comes only as
correction riding on $\Lambda$. In the latter with -ve sign,
there is no Einstein limit and the effective $\lambda$ is always
negative leading to AdS. It stands of its own even when $\Lambda$
vanishes and it can not be switched off. This suggests that its
source is not sitting in the bulk. These two cases clearly
indicate that they refer to two different
situations, and what could they be is what we consider next.  \\
The GB contribution could arise in two different ways. One, when we study the
most general action giving rise to quasi-linear equation for gravitation in
$5-$dimensional spacetime. In this case GB represents
the higher order correction and the Einstein gravity results in the
$\alpha\to0$ limit. This is the case for the +ve sign solution. Second,
when GB term is caused by the second order
iteration of self interaction of gravitational field whose source is sitting
in the $3-$brane. It is purely free gravity
leaking into the bulk from the brane that sources the GB term in the bulk
which can not be switched off in the bulk. This is the case for the -ve sign
solution. As argued earlier, bulk spacetime could either be dS or AdS. Since
it is free gravity that propagates in bulk and has negative energy
density, hence it would generate AdS rather than dS. This demands that the GB
parameter $\alpha$ must be positive to give AdS. We thus end up with a
scenario similar to the Randall-Sundrum braneworld model (RS) \cite{rs} purely
from classical consideration without any reference to string theory. Here
AdS bulk is not an assumption but follows from the property of gravity.
It is therefore no surprise that AdS bulk thus sourced through GB term will
also localize gravity on the brane \cite{rs,gt}, and of course it will have no
$\alpha\to0$ limit in the bulk. These are the two different situations
indicated by the $\pm$ solutions and they get further resolved when a mass
point is introduced. \\
Introduction of a mass point in this setting is
described by the well-known Boulware-Deser solution \cite{bd,jtw} given by
\begin{equation} \label{sol}
ds^2 = - Adt^2 + A^{-1}dr^2 + r^2d\Omega_3^2
\end{equation}
where
\begin{equation} \label{sol1}
A = 1 - \frac{r^2}{2\alpha}[-1 \pm \sqrt{1 + 4\alpha(\frac{M}{r^4} +
\Lambda)}].
\end{equation}
Here $M$ is the mass term which has dimension of $L^2$, and the two solutions
are distinguished by $\pm$ signs. Let us term the +ve sign
solution for which the limit $\alpha \to 0$ exists as the
bulk solution (BS) while the -ve sign one has no
$\alpha \to 0$ limit as the brane-bulk solution (BBS). The term under the 
radical sign must be positive which will be so for $\alpha M>0, 
\alpha\Lambda>0$. For the BS, $M>0$ and consequently $\alpha>0$ will accord 
to the usual attractive gravity in the bulk while it would be repulsive for 
the BBS case unless we reverse the sign of $M, \alpha$. Note that the
metric is nowhere singular and as $r \to 0$ it tends
to $A\to 1 -(\pm \sqrt{\frac{M}{\alpha}}) + \frac{r^2}{2\alpha}$. In the limit
$r=0$, $A \neq 1$ and hence it is not flat but represents a
spacetime asymptotically approximating to a global monopole with a solid 
angle deficit \cite{bv,nyd}. The approach to the limit is however through
AdS. When $M=0$, the limiting space is Minkowski flat. Our main aim is to
probe GB gravity and hence we shall now set $\Lambda =0$, which does not play
any critical role. \\
We define the equivalent Newtonian potential, $\Phi = (A-1)/2$, which leads to
gravitational force given by

\begin{equation}\label{for}
-\Phi^{\prime} = \frac{r}{2\alpha}\left[-1 \pm {\left(1 + 4\alpha \frac{M}
{r^4}\right)^{-1/2}}\right].
\end{equation}

For large $r$ this approximates to the familiar $5-$dimensional Schwarzschild 
for BS while for BBS it is anti-Schwarzshild-AdS unless both 
$M, \alpha$ are -ve, then it would be Schwarzschild-dS. For smaller $r$, it 
goes as $-\frac{r}{2\alpha} \pm O(r^3)$, which shows that approach to the 
centre $r=0$ is always through AdS. This demonstrate the remarkable effect of 
GB
contribution which transforms the radial dependence of gravity, from inverse
to proportional. This is why the singularity structure is radically altered
\cite{torii}. \\

The central singularity is however weak because the Kreschmann scalar
(square of Riemann curvature) diverges only as $r^{-4}$. That means energy
density
will diverge as $r^{-2}$ which on integration over the volume will vanish
as $r\to 0$. This is because at the singularity the metric approximates to
that of a global monopole \cite{nyd} for which this is the characteristic
behavior.
Thus GB contribution, which would be dominant at high energy as singularity is
approached, results in smoothening and weakening the singularity. This is
done not by gravity altering its sense, attraction to repulsion, but by its
behavior transforming from inverse square to proportional to $r$. \\
The BS solution has the Einstein limit $A = 1 - \frac{m^2}{r^2}$, ($M=m^2$)
which is the $5-$dimensional Schwarzschild solution. Note that in
the first approximation, there is no GB contribution and further the higher
order
contribution comes as riding on $M$. It is Minkowski flat when $M=0$,
hence GB contribution has
no existence of its own and it comes only as a riding correction. It has
horizon at $r_h^2 = m^2 - \alpha$ which will exist only if $m^2 \ge \alpha$,
else it will be a naked singularity. Here $\alpha$ behaves like electric 
charge in the Reissner - Nordstr{$\ddot\o$}m solution for a charged black 
hole. Its singularity structure would therefore be similar to it. It is quite 
interesting that 
asymptotically $\alpha$ has no effect while at the horizon it behaves like 
a ``charge''. Though GB contribution comes
in this case only as rider yet its effect becomes dominant as horizon is
reached and it radically changes the horizon and singularity structure
\cite{torii}.\\
The BBS solution with $M = m^2, \alpha>0$ for large $r$ approximates to 
$A = 1 +\frac{m^2}{r^2} + \frac{r^2}{\alpha}$,
which is AS-AdS. Note that the mass point is repulsive. We could however 
reverse the situation by taking $M = -m^2, \alpha<0$, then it would be S-dS. 
Here the GB contribution comes from gravity leaking from
brane into bulk and that produces a spacetime of negative constant curvature,
which is AdS. That is why it will always stand of its own and can not be
switched off unless one switches off gravity entirely in the brane. Clearly,
there is no horizon and there is only weak naked singularity.In this case, 
the background is set up by gravitational field leaking from the 
brane into the bulk which should generate an Ads and hence $\alpha$ must be 
positive. Then addition of a mass point in this setting produces repulsive 
gravity is the most remarkable and intriguing feature which we do not quite 
understand. \\
Our main purpose here was to bring forth and highlight the critical role GB
contribution plays. It is however non-trivial only in $5$ or higher
dimensions. GB gravity arises in two different ways. One, for $n > 4$
dimensions it should be
included in the most general action leading to second order quasi-linear
equation. It is thus a higher order correction which can not stand all by
itself but rides on matter and $\Lambda$ in the higher dimensional spacetime.
On the other hand, GB term could be sourced by free gravity leaking from
$3-$brane into the bulk as second iteration of self interaction. This stands
all by itself and generates an AdS in the bulk. It can not be switched off to
give $\alpha\to 0$ limit simply because its source is not sitting in the
bulk but instead in the brane. The bulk is free of matter and hence it is
maximally symmetric space of constant curvature which is negative because it
is solely produced by free gravitational field having negative energy density.
That is why bulk spacetime has to be an AdS and not dS. Note that it is not
an assumption but follows from the
basic character of gravitational field. On the other hand, in the RS model AdS
bulk is required for localization of gravity \cite{rs,gt}. Further AdS is also 
favoured in a very recent investigation of geodesics and singularities in 
higher dimensional spacetime \cite{et}. Also note that we 
have obtained RS model like scenario purely from classical consideration
without any reference to string theory. We are driven to the
$5-$dimensional bulk simply by the physical realization of second order
iteration of self interaction of gravity. What the second iteration
essentially does is to produce a constant negative curvature in the bulk. A
spacetime of constant curvature however solves the equation (1).  \\
The most interesting case is the BBS where there is a gravitational sharing of
dynamics between brane and bulk. In this case, there never occurs a horizon 
irrespective of whether we have the AS-AdS with $M>0, \alpha>0$ or S-dS with 
$M<0, \alpha<0$. In the braneworld gravity, AdS bulk is required for 
localization of 
gravity on the brane. It has recently been shown that a black hole with
sufficiently large horizon on the bulk will delocalize gravity by sucking in
zero mass gravitons \cite{se}. A mass point in the GB setting presents
a variety of possibilities as there occurs no horizon at all for BBS and
even for BS it could be avoided for $m^2 < \alpha$. The absence of horizon 
altogether in the BBS case is perhaps indicative of the fact that 
localization of gravity on the brane would continue to remain undisturbed by 
the introduction of mass point in the bulk. This is perhaps because our 
interpretation of the BBS is solely guided by the dynamics of gravitational 
field. In this way, GB could therefore play a very important
and interesting role in localizing as well as stabilizing the braneworld
gravity \cite{nd}. \\
The most distinguishing and characteristic feature of the GB gravity is the
negative constant curvature background which manifests as AdS, and its
dominance over the mass at high energy as $r \to 0$ is approached.
Asymptotically as $r\to\infty$, the field goes as $r^{-3}$ for BS and as
AdS + $r^{-3}$ for BBS. At the other end, $r \to 0$, it goes proportional
to $r$. At high energy, gravity effectively changes its radial dependence from
inverse to proportionality. This is what is responsible for smoothening and
weakening of singularity (Similar indication is also emerging when we
consider the dust collapse in the GB setting \cite{sgg}). This makes the
crucial difference in gravitational
dynamics at high and low energy. It is something analogous
to Planck's law of radiation which has similarly different behavior at high
and low energy. In the loop quantum
gravity, apart from gravity turning repulsive there also occurs similar
change at high energy as singularity is approached both in cosmology and
black hole, density transforms from inverse power to positive power of the
scale factor and radius
respectively \cite{mb1,mb2,mb3}. The GB term, which also arises as one loop
contribution in string theory \cite{zw,sen}, seems to anticipate some
aspect of quantum gravity effects at least qualitatively. Thus
it could rightly be considered as intermediate limit of quantum gravity. In
other way, it could be thought of as right pointer to quantum gravity effects.
In the context of loop quantum gravity, we should rather ask for GB gravity
as its intermediate limit and so Ads rather than flat space. That is the
limiting continuum spacetime to loop quantum gravity to be rather
$5-$dimensional AdS than $4-$dimensional flat space. This is the suggestion
which is naturally emerging and hence should deserve serious further
consideration. \\
Very recently there has been an attempt to see a connection between loop 
inspired and braneworld cosmology \cite{cop}. It is shown that the effective 
field equations in the two paradigms bear a dynamical correspondence. 
There appears to be a resonance of it in some other calculation as well 
\cite{par}. Such a bridge between the two approaches to quantum gravity is 
quite expected and most desirable as the two refer to complementary aspects. 
In this perspective, the GB 
term could also be seen as indicative 
of a similar bridge between the two approaches. It is quite rooted in the 
string paradigm through the first loop contribution as well as in the 
braneworld paradigm. It mimics the features similar to that of the loop 
quantum calculations at the high energy regime when singularity is 
approached. Our paradigm makes a very strong suggestion for the intermediate 
semi-classical limit to the loop quantum gravity as AdS $5-$dimensional 
spacetime rather than $4-$dimensional flat spacetime. This is a clear 
prediction. \\
There have been several considerations of higher order terms including GB and 
GB coupled to dilaton in FRW cosmology (see for example \cite{odn,sami}). In 
there, higher order terms act as a matter field in the fixed FRW background 
simply modifying the Friedman equation. It is a prescription while here we 
have a true second order quasi-linear equation to be solved to determine the 
spacetime metric. The two situations are quite different. The former is an 
effective modification of the Einstein's theory while the latter is the 
natural generalization demanded by the dynamics of gravity. \\
It turns out that GB thus has determining say at high energy. 
However all this happens in $5$ dimensions where GB attains
non-trivial physical meaning. It is certainly pointing in the
right direction that quantum gravity effects would at the very
least weaken singularity if not remove it altogether. The most
pertinent question is, could this desirable feature of weakening
of singularity be brought down to $4$ dimensions through dilaton
scalar field coupling to the GB term \cite{des,nn} or otherwise? Very 
recently, a new black hole solution has been found \cite{md} in which 
effects of GB and Kaluza-Klein splitting of spacetime menifest in $4$ 
dimensions. What happens is that GB weakens the singularity and regularizes 
the metric while Kaluza-Klein modes generate the Weyl charge as was the case 
for one of the first black hole solutions on the Randall-Sundrum brane 
described by a charged black hole metric \cite{dmpr}. It is remarkable that 
the new solution asymptotically does indeed approximate to the black hole on 
the brane.

\acknowledgments{I wish to thank Atish Dabholkar and Parampreet
Singh for some clarifying comments and references. }


\end{document}